\documentstyle[a4,amsmath,amssymb,latexsym,epsfig,citesort]{article}
\def \d{{\mathrm{d}}}

\def \pd{\partial}

\def \stimes{\times\!\!\!\!\!\!\supset}

\def \tl#1{\overset{\kern 1pt\circ}{#1}}
\def \TL#1{\overset{\kern -3pt \circ}{#1}}
\def \TLL#1{\overset{\kern -7pt \circ}{#1}}

\begin{document}
\title{{\bf Wedge disclination in the field theory of elastoplasticity}}
\author{Markus Lazar\\
        Max-Planck-Institute for Mathematics in the Sciences,\\
        Inselstr. 22-26, D-04103 Leipzig, Germany\\
        E-mail: lazar@mis.mpg.de}

\date{\today}    
\maketitle
\begin{abstract}
In this paper we study the wedge disclination within the elastoplastic defect theory. 
Using the stress function method we found 
exact analytical solutions for all characteristic fields
of a straight wedge disclination in a cylinder. 
The elastic stress, elastic strain, elastic bend-twist, displacement and rotation 
have no singularities at the disclination line.  
We found a modified stress function for the wedge disclination.
\\

\end{abstract}
\vspace*{8mm}
Disclinations are very important lattice defects analogous to dislocations. 
A disclination is characterized by a closure failure of the rotation
for a closed circuit round the disclination line.
There are wedge and twist disclinations.
It seems that very strong plastic distortions are necessary in order to create disclinations
in crystals.
If the Frank angle (rotation failure) of the disclination is a symmetry angle of lattice, 
then the disclination is called a perfect disclination. 
Such disclinations have been introduced by Anthony~\cite{Anthony70} and deWit~\cite{deWit73a,deWit73b}.
A wedge disclination (rotation axis parallel to the disclination line)
may be generated by adding or removing a wedge-shaped piece of material 
and re-welding the material. 
The smallest value of the Frank vector is 
$\pi/2$ in a cubic lattice and $\pi/3$ in a hexagonal lattice. 
If the Frank angle is not a symmetry angle of lattice, the disclination is 
called partial disclination. 
They play an important role, e.g., in building 
of twin boundaries (see, e.g.,~\cite{deWit72}).
Disclinations correspond, in general, to Volterra's distortions of the second kind 
(see also~\cite{KA75}). 
Thus, these defects are of rotational type. 
They are different from the so-called Frank's (spin) disclinations which are 
(elementary) defects in liquid crystals.

In recent articles~\cite{Lazar02a,Lazar02b,Lazar02c,Lazar02d} we have studied  
screw and edge dislocations in the framework of field theory of elastoplasticity.
This field theory of elastoplasticity can be considered as a gauge theory of defects where
the defects cause plasticity. The corresponding gauge fields may 
be identified with the plastic distortion.
By the help of this theory the elastic and plastic part
of the total distortion can be calculated. 
The total distortion is defined in terms of a displacement and consists of the
elastic and plastic part.
In the case of dislocations~(see, e.g.,~ \cite{Lazar02d})
the elastic distortion is continuous even in the dislocation core
and the plastic part becomes discontinuous.
The field equation of the elastic stress in an isotropic medium is an inhomogeneous Helmholtz equation~(see, e.g.,~\cite{Lazar02c})
\begin{align}
\label{stress-fe}
\Big(1-\kappa^{-2}\Delta\big)\sigma_{ij}=\tl\sigma {}_{ij},\qquad \kappa^2=\frac{2\mu}{a_1}.
\end{align}
Here $\tl\sigma {}_{ij}$ is the classical stress tensor and $\mu$ is the shear modulus.
The coefficient $a_1$ has the dimension of a force and $\kappa$
has the dimension of a reciprocal length.
It is important to note that Eq.~(\ref{stress-fe}) agrees with the 
field equation for the elastic stress in Eringen's nonlocal elasticity~\cite{Eringen83,Eringen87,Eringen2002}
and in gradient elasticity~\cite{GA99a,GA00,Gutkin00}
where the factor $\kappa$ is called non-locality parameter 
or gradient coefficient.
Using the inverse of Hooke's law for $\sigma_{ij}$ and $\tl\sigma {}_{ij}$, 
Eq.~(\ref{stress-fe}) implies an inhomogeneous
Helmholtz equation for the elastic strain 
\begin{align}
\label{strain-fe}
\Big(1-\kappa^{-2}\Delta\big)E_{ij}=\tl E {}_{ij},
\end{align}
where $\tl E {}_{ij}$ is the classical strain tensor.
It worth noting that
Eq.~(\ref{strain-fe}) is analogous to an equation for the elastic strain in 
gradient theory used by Gutkin and Aifantis~\cite{GA99a,GA00,Gutkin00}.
It is, therefore, quite reasonable to use the field theory of elastoplasticity for 
disclinations.

Here we consider a straight wedge disclination inside an infinitely long cylinder with 
outer radius $R$. Disclinations are defined by the disclination line
and the Frank vector.
The $z$-axis is along the disclination line and coincides with the axis of the 
cylinder. 
For a wedge disclination the Frank vector is parallel to the disclination line: 
$\Omega_x=\Omega_y=0$ and $\Omega_z=\Omega$.

In absence of body forces,
the force equilibrium condition can be identically satisfied 
by using the so-called stress function ansatz~\cite{HL}.
If we specialize to the plane problem,
the stress function $f$ is related to the stress tensor 
\begin{align}
\label{stress-ansatz}
\sigma_{ij}=
\left(\begin{array}{ccc}
\pd^2_{yy}f & -\pd^2_{xy}f & 0\\
-\pd^2_{xy}f & \pd^2_{xx}f & 0\\
0& 0& \nu\Delta f
\end{array}\right).
\end{align}
Here $\Delta$ denotes the two-dimensional Laplacian $\pd^2_{xx}+\pd^2_{yy}$
and $\nu$ is Poisson's ratio.
In addition, the strain is given in terms of the stress function as
\begin{align}
\label{strain-ansatz}
E_{ij}=\frac{1}{2\mu}
\left(\begin{array}{ccc}
\pd^2_{yy}f-\nu\Delta f & -\pd^2_{xy}f & 0\\
-\pd^2_{xy}f & \pd^2_{xx}f-\nu\Delta f & 0\\
0& 0& 0
\end{array}\right).
\end{align}
We use the classical stress field of a straight wedge disclination 
in terms of the Airy stress function
\begin{align}
\label{stress-ansatz2}
\tl\sigma {}_{ij}=
\left(\begin{array}{ccc}
\pd^2_{yy}\chi & -\pd^2_{xy}\chi & 0\\
-\pd^2_{xy}\chi & \pd^2_{xx}\chi & 0\\
0& 0& \nu\Delta\chi
\end{array}\right).
\end{align}
The stress function of a ``classical'' wedge disclinations is given by 
\begin{align}
\label{SF-cl}
\chi=A\, r^2\left\{ \ln r -\frac{1-4\nu}{2(1-2\nu)}-C\right\},\qquad A=\frac{\mu \Omega}{4\pi(1-\nu)},
\end{align}
where $r^2=x^2+y^2$. 
That stress function is chosen so that in the case of $C=0$ it reproduces  
the stress of a wedge disclination given by deWit~\cite{deWit73b}.
It fulfils the following inhomogeneous bipotential (or biharmonic) equation 
\begin{align}
\Delta\Delta\,\chi=8\pi A\,\delta(r).
\end{align}
More precisely, the stress function $\chi=\frac{1}{8\pi}\, r^2 \ln r$ is 
Green's function of the two-dimensional bipotential equation. 
Substituting (\ref{stress-ansatz}), (\ref{stress-ansatz2}) and (\ref{SF-cl}) 
into (\ref{stress-fe}) we get 
\begin{align}
\label{f_fe}
\Big(\Delta-\kappa^2\Big)f=-\kappa^2 A\, r^2\left(\ln r -\frac{1-4\nu}{2(1-2\nu)}-C\right).
\end{align}
Now we use the ansatz
\begin{align}
f=A r^2\left(\ln r-\frac{1-4\nu}{2(1-2\nu)}-C\right) +f_{(1)}
\end{align}
and obtain
\begin{align}
\label{f_fe1}
\Big(\Delta-\kappa^2\Big)f_{(1)}=-4A\left(\ln r+\frac{1}{2(1-2\nu)}-C\right).
\end{align}
Its solution reads
\begin{align}
f_{(1)}=\frac{4 A}{\kappa^2}\bigg(\ln r+\frac{1}{2(1-2\nu)}+K_0(\kappa r)-C\bigg).
\end{align}
Finally, we find the solution of (\ref{f_fe})
\begin{align}
\label{f_wedge}
f=\frac{\mu \Omega}{4\pi(1-\nu)}\left\{r^2\bigg(\ln r-\frac{1-4\nu}{2(1-2\nu)}-C\bigg)
+\frac{4}{\kappa^2}\,\bigg(\ln r+K_0(\kappa r)+\frac{1}{2(1-2\nu)}-C\bigg)\right\},
\end{align}
where the first piece is the ``classical'' stress function and $K_n$ is the modified Bessel
function of the second kind and $n=0,1,\ldots$ denotes the order of this function.

\begin{figure}[tp]\unitlength1cm
\vspace*{-1.0cm}
\centerline{
(a)
\begin{picture}(8,6)
\put(0.0,0.2){\epsfig{file=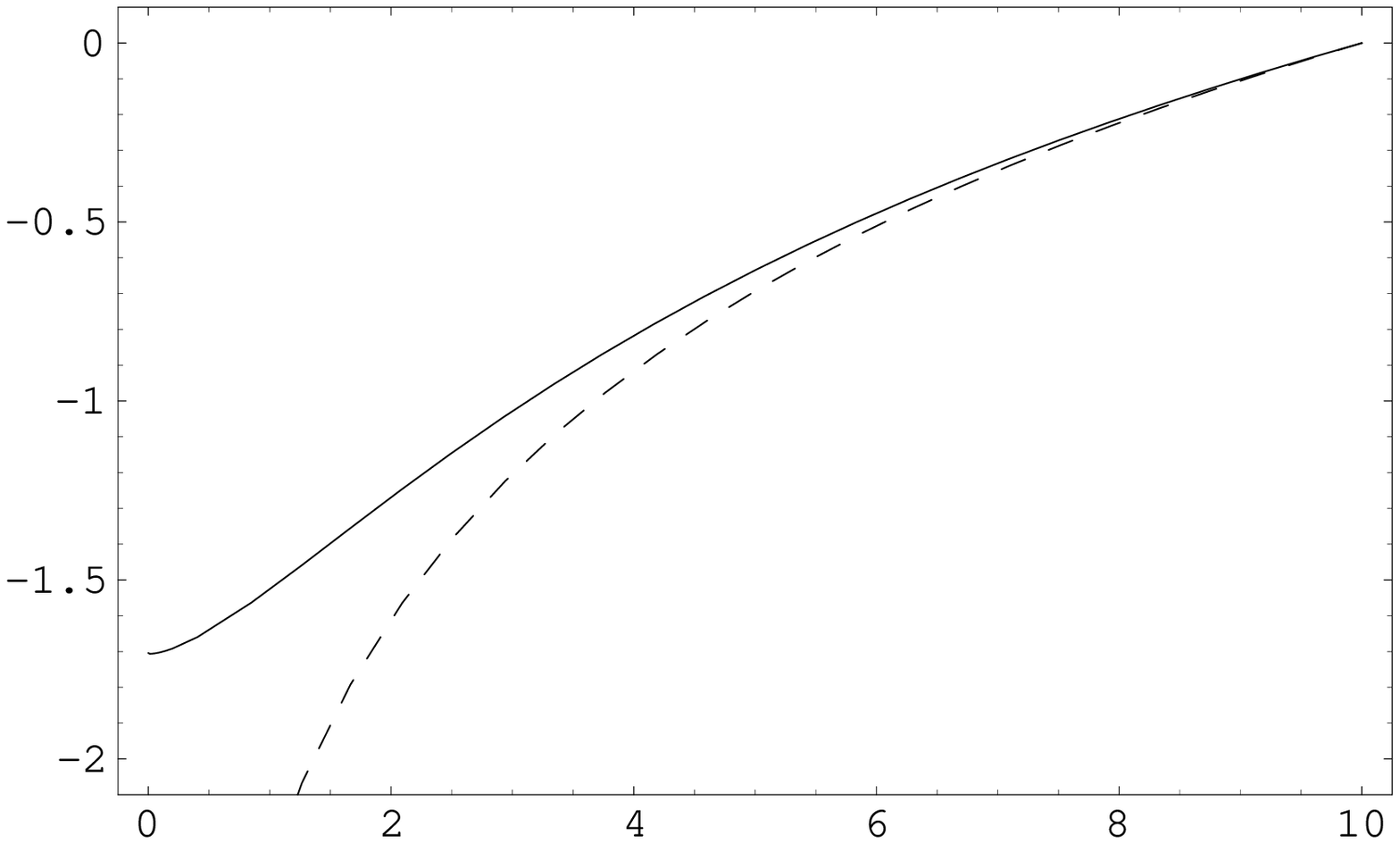,width=9cm}}
\put(4.5,0.0){$\kappa r$}
\put(-1.0,4.5){$\sigma_{rr}$}
\end{picture}
}
\centerline{
(b)
\begin{picture}(8,6)
\put(0.0,0.2){\epsfig{file=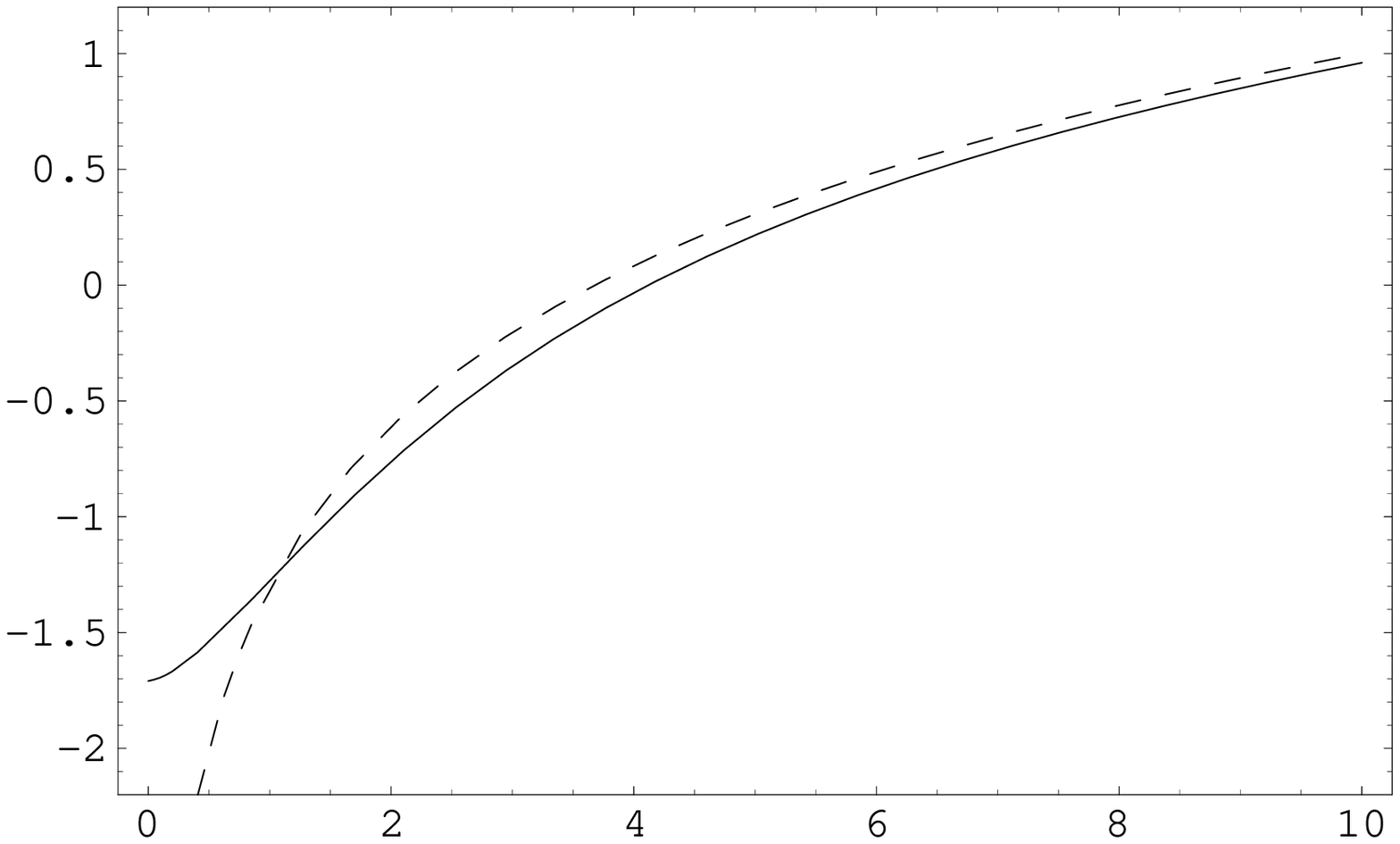,width=9cm}}
\put(4.5,0.0){$\kappa r$}
\put(-1.0,4.5){$\sigma_{\varphi\varphi}$}
\end{picture}
}
\centerline{
(c)
\begin{picture}(8,6)
\put(0.0,0.2){\epsfig{file=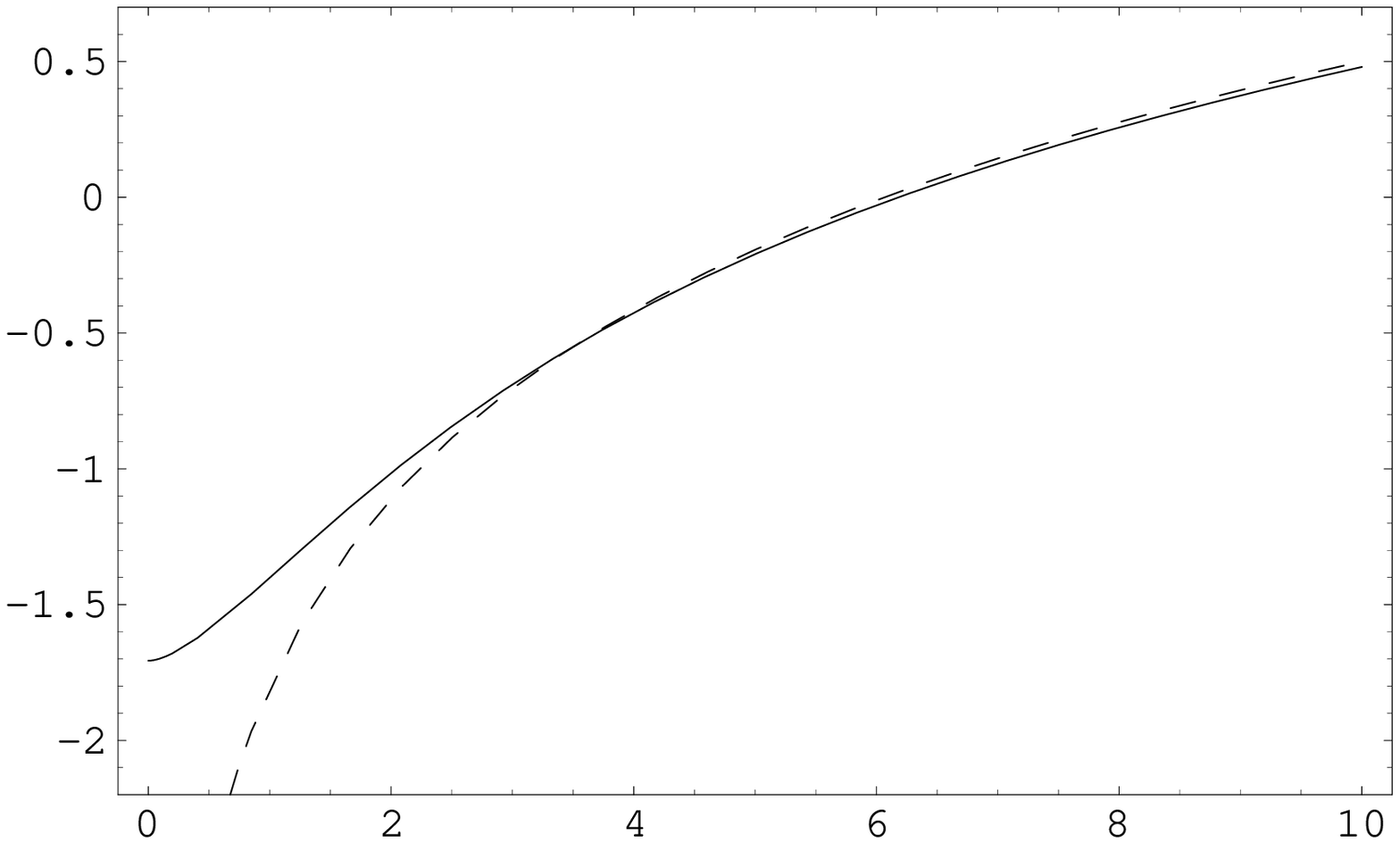,width=9cm}}
\put(4.5,0.0){$\kappa r$}
\put(-1.0,4.4){$\sigma_{zz}$}
\end{picture}
}
\caption{The stress components for a wedge disclination in a 
cylinder with radius $R=10/\kappa$:
(a)~$\sigma_{rr}$, (b)~$\sigma_{\varphi\varphi}$, 
are given in units of $\mu \Omega/[2\pi(1-\nu)]$ 
and 
(c)~$\sigma_{zz}$ is given in units of $\mu \Omega\nu /[\pi(1-\nu)]$. 
The dashed curves represent the classical stress components.}
\label{fig:Trr}
\end{figure}

Let us obtain the stress field of the wedge disclination in the cylinder with radius $R$.
It is convenient to use cylindrical coordinates 
\begin{align}
\sigma_{rr}=\frac{1}{r}\,\pd_r f
,\qquad
\sigma_{\varphi\varphi}=\pd^2_{rr} f
,\qquad
\sigma_{zz}=\nu(\sigma_{rr}+\sigma_{\varphi\varphi}).
\end{align}
In this way we find for the non-vanishing components
\begin{align}
\label{T_rr}
&\sigma_{rr}=\frac{\mu \Omega}{2\pi(1-\nu)}\, 
\bigg\{\ln r+\frac{\nu}{1-2\nu} +K_0(\kappa r)
+\frac{1}{\kappa^2 r^2} \Big(2-\kappa^2 r^2 K_2(\kappa r)\Big)-C\bigg\},\nonumber\\
&\sigma_{\varphi\varphi}=\frac{\mu \Omega}{2\pi(1-\nu)}\, 
\bigg\{\ln r+1+\frac{\nu}{1-2\nu} +K_0(\kappa r)
-\frac{1}{\kappa^2 r^2} \Big(2-\kappa^2 r^2 K_2(\kappa r)\Big)-C\bigg\},\nonumber\\
&\sigma_{zz}=\frac{\mu \Omega\, \nu}{\pi(1-\nu)}\, 
\bigg\{\ln r+\frac{1}{2(1-2\nu)}+K_0(\kappa r)-C\bigg\}.
\end{align}
It can be seen that the boundary term $C$ just gives a constant contribution to the stress~(\ref{T_rr}).
The trace of the stress tensor 
$\sigma_{kk}=\sigma_{rr}+\sigma_{\varphi\varphi}+\sigma_{zz}$ produced by the wedge disclination is
\begin{align}
\label{hyd_p}
\sigma_{kk}=\frac{\mu \Omega(1+\nu)}{\pi(1-\nu)}\, 
\bigg\{\ln r+\frac{1}{2(1-2\nu)}+K_0(\kappa r)-C\bigg\}.
\end{align}
It is interesting to note that for the case when $C=0$ the solution~(\ref{T_rr}) 
agrees with the result obtained by Povstenko~\cite{Pov,Eringen2002} 
in the framework of nonlocal elasticity
if we use the identification $\kappa\equiv 1/(\tau l)$. 
He used the two-dimensional nonlocal 
kernel which is Green's function of the two-dimensional Helmholtz equation.
The constant $C$ is determined by the ``semi-classical'' boundary condition 
on the surface $r=R$ of the cylinder
\begin{align}
\sigma_{rr}(R)=0,
\end{align}
which means the absence of external forces on the boundary of the cylinder.
So we find for the constant
\begin{align}
\label{C}
C= \ln R+\frac{\nu}{1-2\nu} +K_0(\kappa R)
+\frac{1}{\kappa^2 R^2} \Big(2-\kappa^2 R^2 K_2(\kappa R)\Big).
\end{align}
In the limit $\kappa^{-1}\rightarrow 0$, the stress function~(\ref{f_wedge}) 
with (\ref{C}) agrees with the stress function given by~\cite{RV}.
The constant (\ref{C}) diverges in the limit $R\rightarrow\infty$.
Therefore, we consider a cylinder of finite radius.
Note that the cylinder size $R$ is the characteristic screening parameter
which appears under the logarithm.
If we use the limiting expressions for $r\rightarrow 0$,
\begin{align}
\label{K-rel}
K_0(\kappa r)\rightarrow-\left[\gamma+\ln \frac{\kappa r}{2}\right],\qquad
K_2(\kappa r)\rightarrow -\frac{1}{2}+\frac{2}{(\kappa r)^2},
\end{align}
where $\gamma$ denotes the Euler constant,
we obtain 
\begin{align}
\label{Trr(0)}
&\sigma_{rr}(0)=\sigma_{\varphi\varphi}(0)=\frac{1}{2\nu}\,\sigma_{zz}(0)=
\frac{1}{2(1+\nu)}\,\sigma_{kk}(0)=
-\frac{\mu \Omega}{2\pi(1-\nu)}\, 
\bigg\{\ln \frac{\kappa R}{2}+\gamma-\frac{1}{2}+K_0(\kappa R)\nonumber\\
&\hspace{8cm}
+\frac{1}{\kappa^2 R^2} \Big(2-\kappa^2 R^2 K_2(\kappa R)\Big)\bigg\}.
\end{align}
Consequently, the stress is finite at the disclination line in contrast to 
the unphysical stress singularity
in ``classical'' disclination theory. 
The graphs of the components of the stress calculated from~(\ref{T_rr}) 
are plotted over $\kappa r$ in Fig.~\ref{fig:Trr}
(with the radius of the cylinder $R=10/\kappa$). 
With $R=10/\kappa$ we obtain for~(\ref{Trr(0)}) the value
\begin{align}
\sigma_{rr}(0)=\sigma_{\varphi\varphi}(0)=\frac{1}{2\nu}\,\sigma_{zz}(0)=
\frac{1}{2(1+\nu)}\,\sigma_{kk}(0)
\simeq -1.707\,\frac{\mu \Omega}{2\pi(1-\nu)}.
\end{align}
At the outer boundary $R$ the non-vanishing components of the stress 
tensor~(\ref{T_rr}) are
\begin{align}
\label{Tpp(R)}
\sigma_{\varphi\varphi}(R)=\frac{1}{\nu}\,\sigma_{zz}(R)=
\frac{1}{(1+\nu)}\,\sigma_{kk}(R)=
\frac{\mu \Omega}{2\pi(1-\nu)}\, 
\bigg\{1-\frac{4}{\kappa^2 R^2}+2 K_2(\kappa R)\bigg\}.
\end{align}
With $R=10/\kappa$ we obtain for~(\ref{Tpp(R)}) the value (see Fig.~\ref{fig:Trr})
\begin{align}
\sigma_{\varphi\varphi}(R)=\frac{1}{\nu}\,\sigma_{zz}(R)=
\frac{1}{(1+\nu)}\,\sigma_{kk}(R)
\simeq 0.960\,\frac{\mu \Omega}{2\pi(1-\nu)}.
\end{align}

By means of Eqs.~(\ref{stress-ansatz}) and (\ref{f_wedge}) we obtain 
the modified stress of the wedge disclination in the cylinder with radius $R$
in Cartesian coordinates
\begin{align}
\label{T_xx}
&\sigma_{xx}=\frac{\mu \Omega}{2\pi(1-\nu)}\, 
\bigg\{\ln r+\frac{y^2}{r^2}+\frac{\nu}{1-2\nu} +K_0(\kappa r)
+\frac{\big(x^2-y^2\big)}{\kappa^2 r^4} \Big(2-\kappa^2 r^2 K_2(\kappa r)\Big)-C\bigg\},\nonumber\\
&\sigma_{yy}=\frac{\mu \Omega}{2\pi(1-\nu)}\, 
\bigg\{\ln r+\frac{x^2}{r^2}+\frac{\nu}{1-2\nu} +K_0(\kappa r)
-\frac{\big(x^2-y^2\big)}{\kappa^2 r^4} \Big(2-\kappa^2 r^2 K_2(\kappa r)\Big)-C\bigg\},\nonumber\\
&\sigma_{xy}=-\frac{\mu \Omega}{2\pi(1-\nu)}\, 
\frac{xy}{r^2}\bigg\{1-\frac{2}{\kappa^2 r^2} \Big(2-\kappa^2 r^2 K_2(\kappa r)\Big)\bigg\},\nonumber\\
&\sigma_{zz}=\frac{\mu \Omega\, \nu}{\pi(1-\nu)}\, 
\bigg\{\ln r+\frac{1}{2(1-2\nu)}+K_0(\kappa r)-C\bigg\}.
\end{align}
If we identify $\kappa\equiv 1/\sqrt{c}$ ($c$ is the gradient coefficient
used by Gutkin and Aifantis) and put $C=0$, the elastic stress is in agreement with the stress field 
given by Gutkin and Aifantis~\cite{GA00,Gutkin00}
in the framework of strain gradient elasticity by using the Fourier transform
method.

Using (\ref{strain-ansatz}) and (\ref{f_wedge})
we find for the elastic strain of the straight wedge disclination 
\begin{align}
\label{strain_cart}
&E_{xx}=\frac{\Omega}{4\pi(1-\nu)}
\bigg\{ (1-2\nu)\big(\ln r +K_0(\kappa r)-C\big)
+\frac{y^2}{r^2}+\frac{(x^2-y^2)}{\kappa^2 r^4}\Big(2-\kappa^2 r^2 K_2(\kappa r)\Big)\bigg\},\nonumber\\
&E_{yy}=\frac{\Omega}{4\pi(1-\nu)}
\bigg\{ (1-2\nu)\big(\ln r +K_0(\kappa r)-C\big)
+\frac{x^2}{r^2}-\frac{(x^2-y^2)}{\kappa^2 r^4}\Big(2-\kappa^2 r^2 K_2(\kappa r)\Big)\bigg\},\nonumber\\
&E_{xy}=-\frac{\Omega}{4\pi(1-\nu)}\, 
\frac{xy}{r^2}\bigg\{1-\frac{2}{\kappa^2 r^2} \Big(2-\kappa^2 r^2 K_2(\kappa r)\Big)\bigg\}.
\end{align}
In the case of $C=0$
the strain~(\ref{strain_cart}) coincides with the result given by Gutkin and Aifantis~\cite{GA00,Gutkin00,GA99}.
The strain reads in cylindrical coordinates
\begin{align}
\label{E-polar}
&E_{rr}=\frac{\Omega}{4\pi(1-\nu)}
\bigg\{ (1-2\nu)\big(\ln r +K_0(\kappa r)-C\big)
+\frac{1}{\kappa^2 r^2}\Big(2-\kappa^2 r^2 K_2(\kappa r)\Big)\bigg\},\nonumber\\
&E_{\varphi\varphi}=\frac{\Omega}{4\pi(1-\nu)}
\bigg\{ (1-2\nu)\big(\ln r +K_0(\kappa r)-C\big)+1
-\frac{1}{\kappa^2 r^2}\Big(2-\kappa^2 r^2 K_2(\kappa r)\Big)\bigg\}.
\end{align}
The dilatation reads
\begin{align}
\label{dilat}
E_{kk}=\frac{\Omega}{2\pi(1-\nu)}
\bigg\{(1-2\nu)\big(\ln r+K_0(\kappa r)-C\big)+\frac{1}{2}\bigg\}.
\end{align}
By the help of~(\ref{K-rel}) 
the expressions of the strain~(\ref{E-polar}) and the dilatation~(\ref{dilat}) 
read for the limit $r\rightarrow 0$, respectively, 
\begin{align}
\label{Err(0)}
&E_{rr}(0)=E_{\varphi\varphi}(0)=
-\frac{\Omega}{4\pi(1-\nu)}\, 
\bigg\{
(1-2\nu)\left[
\ln \frac{\kappa R}{2}+\gamma-\frac{1}{2}+K_0(\kappa R)\right]\nonumber\\
&\hspace{8cm}
+\frac{1}{\kappa^2 R^2} \Big(2-\kappa^2 R^2 K_2(\kappa R)\Big)\bigg\},
\end{align}
and
\begin{align}
\label{Ekk(0)}
E_{kk}(0)=-\frac{\Omega(1-2\nu)}{2\pi(1-\nu)}\, 
\bigg\{\ln \frac{\kappa R}{2}+\gamma-\frac{1}{2}+K_0(\kappa R)
+\frac{1}{\kappa^2 R^2} \Big(2-\kappa^2 R^2 K_2(\kappa R)\Big)\bigg\}.
\end{align}
The strain and the dilatation at the cylindrical surface $r=R$ are given by
\begin{align}
\label{Epp(R)}
E_{\varphi\varphi}(R)=
-\frac{1-\nu}{\nu}\,E_{rr}(R)=
\frac{1-\nu}{(1-2\nu)}\, E_{kk}(R)=
\frac{\Omega}{4\pi}\, 
\bigg\{1-\frac{4}{\kappa^2 R^2}+2 K_2(\kappa R)\bigg\}.
\end{align}

\begin{figure}[t]\unitlength1cm
\centerline{
\begin{picture}(9,6)
\put(0.0,0.2){\epsfig{file=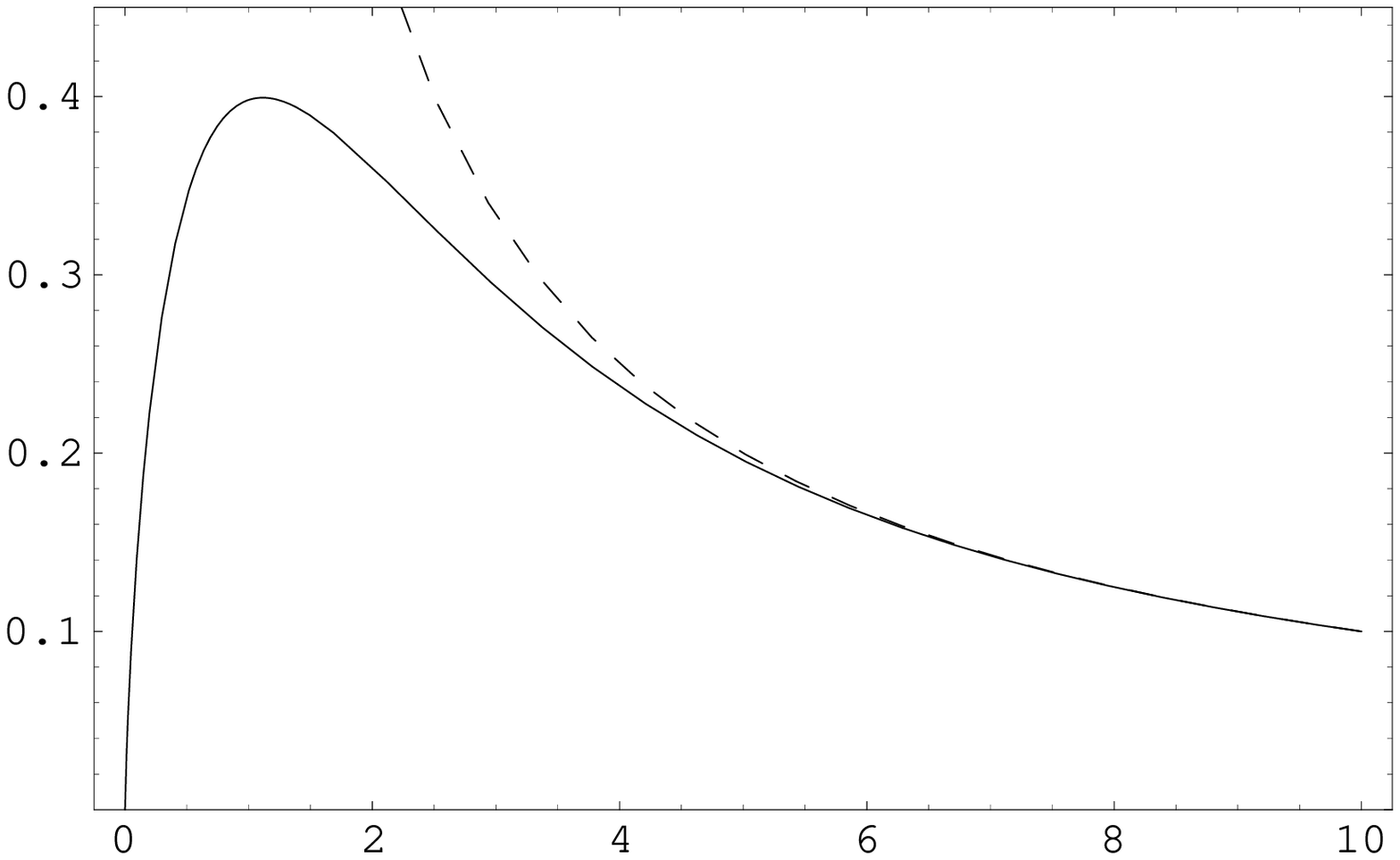,width=9cm}}
\put(4.5,0.0){$\kappa r$}
\put(-1.0,4.5){$k_{z\varphi}$}
\end{picture}
}
\caption{Elastic bend-twist $k_{z\varphi}$ of a wedge disclination (solid)
is given in units of $\Omega/[2\pi]$.
The dashed curve represents the classical solution.}
\label{fig:k_zp}
\end{figure}

In the conventional disclination theory the torsion tensor (linear version of Cartan's torsion)
is defined by
\begin{align}
\label{tors1}
\alpha_{ij}=
\epsilon_{jkl}\big(\pd_k\beta_{il}+\epsilon_{ilm}\varphi^*_{mk}\big).
\end{align}
It is the dislocation density in the theory of disclinations 
(see, e.g., \cite{Mura72,deWit73a,deWit73b}). 
On the other hand, Anthony~\cite{Anthony70}
called it the disclination torsion.
$\beta_{ij}$ is the elastic distortion.
The $\varphi^*_{ij}$ was introduced by Mura~\cite{Mura72}
who called it ``plastic rotation'' and deWit~\cite{deWit73a,deWit73b,deWit72b} called
this quantity ``disclination loop density''.
Using the elastic  bend-twist tensor (see, e.g.,~\cite{Kosseka})
\begin{align}
\label{bt}
k_{ij}=\pd_j\omega_i-\varphi^*_{ij},
\end{align}
with the rotation vector 
\begin{align}
\omega_i=-\frac{1}{2}\,\epsilon_{ijk}\beta_{jk},
\end{align}
Eq.~(\ref{tors1}) can be rewritten according to (see also~\cite{Anthony70,deWit72b,Kosseka})  
\begin{align}
\alpha_{ij}=
\epsilon_{jkl}\big(\pd_k E_{il}+\epsilon_{iml} k_{mk}\big)
=\epsilon_{jkl}\pd_k E_{il}+\delta_{ij} k_{ll}-k_{ji}.
\end{align}
Consequently, the elastic bend-twist  may be determined from 
the condition
that the dislocation density (disclination torsion) has to be zero for a straight wedge
disclination
\begin{align}
\label{dd2}
\alpha_{xz}=-\frac{1-\nu}{2\mu}\,\pd_y\Delta f -k_{zx}\equiv 0,\qquad
\alpha_{yz}=\frac{1-\nu}{2\mu}\,\pd_x\Delta f-k_{zy}\equiv 0.
\end{align}
So we find for the elastic bend-twist
\begin{align}
\label{bt-el}
k_{zx}=-\frac{\Omega}{2\pi}\,
\frac{y}{r^2}\Big\{1-\kappa r K_1(\kappa r)\Big\},\qquad
k_{zy}=\frac{\Omega}{2\pi}\,
\frac{x}{r^2}\Big\{1-\kappa r K_1(\kappa r)\Big\},
\end{align}
and in cylindrical coordinates
\begin{align}
\label{k_zp}
k_{z\varphi}=\frac{\Omega}{2\pi}\,
\frac{1}{r}\Big\{1-\kappa r K_1(\kappa r)\Big\},\qquad 
k_{zr}=0.
\end{align}
Obviously, the components of the elastic bend-twist~(\ref{bt-el}) and (\ref{k_zp})
fulfil an inhomogeneous Helmholtz equation
\begin{align}
\label{k-fe}
\Big(1-\kappa^{-2}\Delta\big)k_{ij}=\tl k {}_{ij},
\end{align}
where $\tl k {}_{ij}$ denotes the classical elastic bend-twist.
The elastic bend-twist $k_{z\varphi}$ is plotted in Fig.~\ref{fig:k_zp}. 
It has no singularity at the disclination line. It is $k_{z\varphi}(r=0)=0$.
The elastic bend-twist
has a maximum of 
$k^{\mathrm{max}}_{z\varphi}\simeq 0.399\kappa \Omega/[2\pi]$ at $r\simeq 1.1 /\kappa$. 
The boundary condition $k_{zr}(R)=0$ is trivially satisfied.

\begin{figure}[t]\unitlength1cm
\centerline{
\begin{picture}(9,6)
\put(0.0,0.2){\epsfig{file=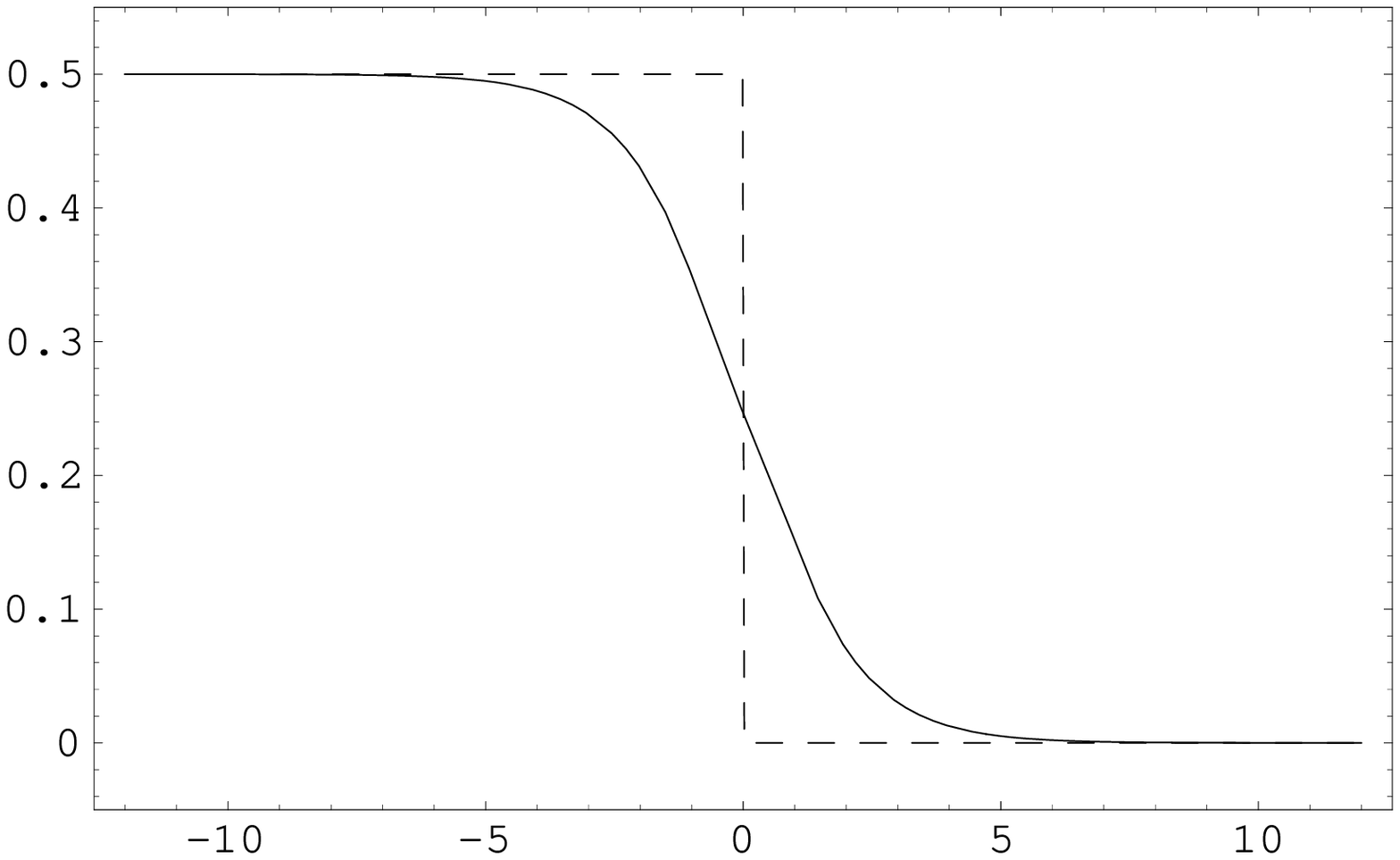,width=9cm}}
\put(4.5,0.0){$\kappa x$}
\put(-2.0,4.5){$\omega_z(x,0)/\Omega$}
\end{picture}
}
\caption{Rotation vector of a wedge disclination $\omega_z(x,y\rightarrow +0)$ 
(solid).
The dashed curve represents the classical solution.}
\label{fig:rot}
\end{figure}

The elastic bend-twist tensor can be decomposed according to~(\ref{bt})
into a gradient of the rotation vector and an incompatible part.
We identify the incompatible part with the disclination loop density.
It is analogous to the decomposition of the elastic distortion of a dislocation
into a gradient of the displacement vector and an incompatible distortion 
(see~\cite{Lazar02a,Lazar02b,Lazar02c}).
The disclination loop density turns out to be 
\begin{align}
\label{bt-pl}
&\varphi^*_{zx}=\frac{\Omega}{2\pi}\,
\kappa^2 x K_0(\kappa r)\Big(\varphi-\frac{\pi}{2}\,{\mathrm{sign}}(y)\Big),\nonumber\\
&\varphi^*_{zy}=\frac{\Omega}{2\pi}\Big\{
\kappa^2 y K_0(\kappa r)\Big(\varphi
-\frac{\pi}{2}\,{\mathrm{sign}}(y)\Big)
+\pi\delta(y) \Big(1-{\mathrm{sign}}(x)\big[1-\kappa r K_1(\kappa r)\big]\Big)\Big\}.
\end{align}
It contains the angle $\varphi$ and its form is analogous to the plastic distortion 
of a screw dislocation~(see~\cite{Lazar02c}).
Here we use a single-valued discontinuous form of $\varphi$ 
(see~\cite{deWit73b}).
Only the component $\varphi^*_{zy}$ has a $\delta$-singularity 
at $y=0$ like the disclination loop density~\cite{Mura72,deWit73b} 
$\varphi^*_{yz}=(\Omega/2)\,\delta(y)(1-{\mathrm{sign}}(x))$.
The rotation vector reads
\begin{align}
\label{rot-vec}
\omega_z=
\frac{\Omega}{2\pi}\,\Big\{\varphi\big(1-\kappa r K_1(\kappa r)\big)
+\frac{\pi}{2}\,{\mathrm{sign}}(y)\,\kappa r K_1(\kappa r)\Big\}.
\end{align}
The far field of the rotation vector~(\ref{rot-vec}) 
agrees with deWit's expression given in~\cite{deWit73b}. 
When $y\rightarrow +0$, the expression~(\ref{rot-vec}) is plotted
in Fig.~\ref{fig:rot}. One can see that the Bessel function terms
which appear in~(\ref{rot-vec}) lead to the symmetric smoothing
of the rotation vector profile, in contrast to the abrupt jump
occurring in the classical solution. It is interesting to note that the size
of such a transition zone is approximately $12/\kappa$ which gives 
the value $6/\kappa$ for the radius of the disclination core. 
Replacing $\Omega$ by $b$ in Eq.~(\ref{rot-vec}) it can be seen that
the rotation of the wedge disclination $\omega_z$ 
has the same form as the displacement of a screw dislocation $u_z$
given in~\cite{Lazar02c}.
Consequently, the elastic rotation (antisymmetric part of the elastic distortion) 
of a wedge disclination is discontinuous due to $\varphi$ 
in contrast to the dislocation case where the corresponding expressions
are continuous (see also~\cite{deWit73b}). 

Finally, we find for the elastic distortion 
\begin{align}
\label{dist-el}
&\beta_{xx}=\frac{\Omega}{4\pi(1-\nu)}
\bigg\{ (1-2\nu)\big(\ln r +K_0(\kappa r)-C\big)
+\frac{y^2}{r^2}+\frac{(x^2-y^2)}{\kappa^2 r^4}\Big(2-\kappa^2 r^2 K_2(\kappa r)\Big)\bigg\},\nonumber\\
&\beta_{xy}=-\frac{\Omega}{4\pi(1-\nu)}\, 
\frac{xy}{r^2}\bigg\{1-\frac{2}{\kappa^2 r^2} \Big(2-\kappa^2 r^2 K_2(\kappa r)\Big)\bigg\}
\nonumber\\
&\hspace{5cm}
-\frac{\Omega}{2\pi}\,\Big\{\varphi\big(1-\kappa r K_1(\kappa r)\big)
+\frac{\pi}{2}\,{\mathrm{sign}}(y)\,\kappa r K_1(\kappa r)\Big\},\nonumber\\
&\beta_{yx}=-\frac{\Omega}{4\pi(1-\nu)}\, 
\frac{xy}{r^2}\bigg\{1-\frac{2}{\kappa^2 r^2} \Big(2-\kappa^2 r^2 K_2(\kappa r)\Big)\bigg\}
\nonumber\\
&\hspace{5cm}
+\frac{\Omega}{2\pi}\,
\Big\{\varphi\big(1-\kappa r K_1(\kappa r)\big)
+\frac{\pi}{2}\,{\mathrm{sign}}(y)\,\kappa r K_1(\kappa r)\Big\},\nonumber\\
&\beta_{yy}=\frac{\Omega}{4\pi(1-\nu)}
\bigg\{ (1-2\nu)\big(\ln r +K_0(\kappa r)-C\big)
+\frac{x^2}{r^2}-\frac{(x^2-y^2)}{\kappa^2 r^4}\Big(2-\kappa^2 r^2 K_2(\kappa r)\Big)\bigg\}.
\end{align}
We note that the elastic distortion~(\ref{dist-el}) contains 
the angle $\varphi$ in contrast to the dislocation case.
The elastic distortion satisfies the following relations 
that the Burgers vector vanishes for a wedge disclination
\begin{align}
\oint_\gamma\big(\beta_{xx}\d x+\beta_{xy}\d y\big)=0,\qquad
\oint_\gamma\big(\beta_{yx}\d x+\beta_{yy}\d y\big)=0.
\end{align}
Here $\gamma$ denotes the Burgers circuit.

We obtain for the effective Frank vector of the wedge disclination
\begin{align}
\label{Frank-eff}
\Omega(r)=\oint_\gamma \big(k_{zx}\d x+k_{zy}\d y\big)
        =\Omega\,\Big\{1-\kappa r K_1(\kappa r)\Big\}.
\end{align}
This expression is in agreement with the one obtained in~\cite{VS88}.
The form of the effective Frank vector is similar to the effective Burgers vector
of screw and edge dislocations (see, e.g.,~\cite{Lazar02a,Lazar02b,Lazar02c,Lazar02d}).
It differs appreciably from the constant value $\Omega$ in the 
region from $r=0$ up to $r\simeq 6/\kappa$. 
In fact, we find $\Omega(0)=0$ and $\Omega(\infty)=\Omega$.
Thus, it is suggestive to
take $r_c\simeq 6/\kappa$ as the core radius of the disclination.

The so-called disclination density tensor of a discrete disclination
is defined by~\cite{Anthony70,deWit73a,deWit73b,Mura72,deWit72b,Kosseka}
\begin{align}
\Theta_{ij}\equiv\epsilon_{jmn}\pd_m k_{in}=-\epsilon_{jmn}\pd_m \varphi^*_{in}.
\end{align}
Here the index $i$ indicates the direction of the Frank vector, 
$j$ the disclination line direction. 
Thus, the diagonal components of $\Theta_{ij}$ represent wedge disclinations,
the off-diagonal components twist disclinations.
The disclination density tensor satisfies the continuity condition $\pd_j\Theta_{ij}=0$
which implies that disclinations do not end inside the body. 
From (\ref{bt-el}) and (\ref{bt-pl})
we find for the non-vanishing component of the disclination density of a wedge
disclination
\begin{align}
\label{den-wedge}
\Theta_{zz}=\frac{\Omega\kappa^2}{2\pi}\, K_0(\kappa r).
\end{align}
In the limit as $\kappa^{-1}\rightarrow0$, 
the result~(\ref{den-wedge}) converts to the classical expression
$\Theta_{zz}=\Omega\, \delta(r)$.
It is interesting to note that (\ref{den-wedge}) coincides with the 
expression calculated in~\cite{VS88}. 
Additionally, 
the disclination density~(\ref{den-wedge}) agrees with Eringen's 
two-dimensional nonlocal kernel used in~\cite{Eringen83,Eringen87,Eringen2002}.

We use the decomposition of the elastic distortion~(\ref{dist-el}) into a compatible and
an incompatible distortion
\begin{align}
\beta_{ij}=\pd_j u_i+\tilde\beta_{ij}.
\end{align}
In this way we restore a modified displacement vector 
(see also~\cite{Lazar02c})
\begin{align}
\label{displ}
u_x&=-\frac{\Omega}{2\pi}\bigg\{
        y\Big[\varphi\big(1-\kappa r K_1(\kappa r)\big)
        +\frac{\pi}{2}\,{\mathrm{sign}}(y)\,\kappa r K_1(\kappa r)\Big]\nonumber\\
        &\qquad\qquad-\frac{x}{2(1-\nu)}\Big[(1-2\nu)\big(\ln r -1 +K_0(\kappa r)-C\big)
        -\frac{1}{\kappa^2 r^2}\,\big(2-\kappa^2 r^2 K_2(\kappa r)\big)\Big]\bigg\},\nonumber\\
u_y&=\frac{\Omega}{2\pi}\bigg\{
        x\Big[\varphi\big(1-\kappa r K_1(\kappa r)\big)
        +\frac{\pi}{2}\,{\mathrm{sign}}(y)\,\kappa r K_1(\kappa r)\Big]\nonumber\\
        &\qquad\qquad
        +\frac{y}{2(1-\nu)}\Big[(1-2\nu)\big(\ln r -1 +K_0(\kappa r)-C\big)
        -\frac{1}{\kappa^2 r^2}\,\big(2-\kappa^2 r^2 K_2(\kappa r)\big)\Big]\bigg\},
\end{align}
and the incompatible part 
\begin{align}
\label{phi}
&\tilde\beta_{xx}=\frac{\Omega}{2\pi}\,\Big\{\kappa r K_1(\kappa r)
+\kappa^2 x y K_0(\kappa r)
\Big(\varphi-\frac{\pi}{2}\,{\mathrm{sign}}(y)\Big)\Big\},\nonumber\\
&\tilde\beta_{xy}=\frac{\Omega}{2\pi}\, \kappa^2 y^2 K_0(\kappa r)
\Big(\varphi-\frac{\pi}{2}\,{\mathrm{sign}}(y)\Big)
,\nonumber \\
&\tilde\beta_{yx}=-\frac{\Omega}{2\pi}\, \kappa^2 x^2 K_0(\kappa r)
\Big(\varphi-\frac{\pi}{2}\,{\mathrm{sign}}(y)\Big),\\
&\tilde\beta_{yy}=\frac{\Omega}{2\pi}\,\Big\{\kappa r K_1(\kappa r)
-\kappa^2 x y K_0(\kappa r)
\Big(\varphi-\frac{\pi}{2}\,{\mathrm{sign}}(y)\Big)
-\pi\delta(y)\, x \Big(1-{\mathrm{sign}}(x)\big[1-\kappa r K_1(\kappa r)\big]\Big)\Big\}.\nonumber
\end{align}
Eq.~(\ref{phi}) fulfils $\alpha_{ij}=\epsilon_{jkl}(\pd_k\tilde\beta_{il}+\epsilon_{ilm}\varphi^*_{mk})\equiv 0$
and Eqs.~(\ref{displ}) and (\ref{phi}) satisfy 
$\omega=\pd_{[y} u_{x]}+\tilde\beta_{[xy]}$.
The $\delta$-term in~(\ref{phi}) has a similar form like 
the plastic strain or dislocation loop density
of a wedge disclination~\cite{Mura72,deWit72b,deWit73b}
$E^P_{yy}\equiv\beta^*_{yy}=-(\Omega/2)\,\delta(y)x(1-{\mathrm{sign}}(x))$.

Using the gauge theory of defect~\cite{EL88}, which is an $ISO(3)$-gauge theory
($ISO(3)=T(3)\stimes SO(3)$),  
we are able to decompose the incompatible distortion~(\ref{phi}).
Namely, the incompatible distortion takes the form~\cite{EL88,Maly}
\begin{align}
\tilde\beta_{ij}=\phi_{ij}+\epsilon_{ikl} W_{kj} x_l,
\end{align}
where $\phi_{ij}$ and $W_{ij}$ are the translational  and rotational gauge 
fields, respectively. 
We obtain for~(\ref{phi}) the following decomposition
\begin{align}
&\tilde\beta_{xx}=\phi_{xx}-y W_{zx},\nonumber\\
&\tilde\beta_{xy}=-y W_{zy},\nonumber\\
&\tilde\beta_{yx}=x W_{zx},\nonumber\\
&\tilde\beta_{yy}=\phi_{yy}+x W_{zy},
\end{align}
into the translational gauge field 
\begin{align}
\phi_{xx}=\phi_{yy}=\frac{\Omega}{2\pi}\,\kappa r K_1(\kappa r),
\end{align}
and the rotational gauge field
\begin{align}
\label{W-pot}
W_{zx}\equiv-\varphi^*_{zx},\qquad W_{zy}\equiv-\varphi^*_{zy}.
\end{align}
Therefore, the negative disclination loop density~(\ref{bt-pl}) 
is equivalent to the rotational gauge potential~(\ref{W-pot}).

In conclusion, the field theory of elastoplasticity has been employed on the consideration
of a straight wedge disclination. 
We were able to calculate the elastic and plastic fields.
We found that the elastic stress, elastic strain, elastic bend-twist 
and disclination density are 
continuous and the displacement, plastic distortion, rotation and 
the disclination loop density
of the wedge disclination are discontinuous fields.
Exact analytical solutions 
for all characteristic field quantities of a wedge disclination in a cylinder have been 
reported which demonstrate the elimination of ``classical'' logarithmic singularities 
at the disclination line. 
All logarithmic terms are influenced by the ``semi-classical'' boundary term $C$~(\ref{C}).
In addition, the disclination core appears naturally as a result of 
the smoothing of the rotation vector profile.
For an infinitely extended body ($C=0$)
the elastic stress of a wedge disclination calculated in the field theory of 
elastoplasticity agrees with the stress calculated within
the theory of nonlocal elasticity and strain gradient elasticity. 
The reason is that in all three theories the fundamental equation
for the elastic stress has the form of an inhomogeneous Helmholtz
equation (see Eq.~(\ref{stress-fe})).
The boundary-value problem of a wedge disclination in a cylinder 
considered in this paper should be help in studies of
mechanical behaviour of nano-objects including
nanotubes and nanomembranes and of disclinated nanoparticles
of cylindrical shape (nanowires).
Finally, we note that one observes an interesting relation between the wedge disclination
and the screw dislocation.
Namely the rotation~(\ref{rot-vec}), elastic bend-twist~(\ref{k_zp}), 
effective Frank vector~(\ref{Frank-eff})
and the disclination density~(\ref{den-wedge}) of a wedge disclination 
have the same form as the displacement, elastic distortion, effective Burgers vector
and the dislocation density of a screw dislocation given in~\cite{Lazar02b,Lazar02c} 
when the Frank vector is replaced by the Burgers vector.

\subsection*{Acknowledgement}
The author is very grateful to Dr.~Mikhail Yu.~Gutkin for  
useful comments and proposals on an earlier version of this paper.
He acknowledges the Max-Planck-Institut f{\"u}r 
Mathematik in den Naturwissenschaften for financial support.


\begin{thebibliography}{99}
\bibitem{Anthony70} K.-H.~Anthony, Arch. Rat. Mech. Anal.~{\bf 39} (1970) 43.
\bibitem{deWit73a} R.~deWit, 
        J.~Res.~Nat. Bur. Stand. (U.S.)~{\bf 77A} (1973) 49.
\bibitem{deWit73b} R.~deWit, 
        J.~Res.~Nat. Bur. Stand. (U.S.)~{\bf 77A} (1973) 607.
\bibitem{deWit72} R.~deWit, J. Phys. C: Solid State Phys.~{\bf 5} (1972) 529.
\bibitem{KA75} E.~Kr{\"o}ner and K.-H.~Anthony, Ann. Rev. Mat.~{\bf 5} (1975) 43.
\bibitem{Lazar02a} M.~Lazar, 
        J. Phys. A: Math. Gen.~{\bf 35} (2002) 1983.
\bibitem{Lazar02b} M.~Lazar, 
        Ann. Phys.~(Leipzig)~{\bf 11} (2002) 635.
\bibitem{Lazar02c} M.~Lazar, 
        J.~Phys.~A: Math. Gen.~{\bf 36} (2003) 1415.
\bibitem{Lazar02d} M.~Lazar, 
    {\it Dislocations in the Field Theory of Elastoplasticity}, 
        to appear in: Computational Materials Science (2003).
\bibitem{Eringen83} A.C.~Eringen, J. Appl. Phys.~{\bf 54} (1983) 4703.
\bibitem{Eringen87} A.C.~Eringen, 
        Res. Mech.~{\bf 21} (1987) 313.
\bibitem{Eringen2002} A.C.~Eringen, {\it Nonlocal Continuum Field Theories},
        Springer, New York (2002).
\bibitem{GA99a} M.Yu.~Gutkin and E.C.~Aifantis, 
        Scripta Mater.~{\bf 40} (1999) 559.
\bibitem{GA00} M.Yu.~Gutkin and E.C.~Aifantis, in:
        {\it Nanostructured Film and Coatings},
        NATO ARW Series, High Technology, Vol. 78, ed. by G.M.~Chow et al. 
        (Kluwer, Dodrecht, 2000) p.~247.
\bibitem{Gutkin00} M.Yu.~Gutkin, Rev. Adv. Mater. Sci.~{\bf 1} (2000) 27.
\bibitem{HL} J.P.~Hirth and J.~Lothe, {\it Theory of Dislocations}, 
        2nd edition, John Wiley, New York (1982).
\bibitem{Pov} Yu.Z.~Povstenko, Int. J. Engng. Sci.~{\bf 33} (1995) 575.
\bibitem{RV} A.E.~Romanov and V.I.~Vladimirov, {\it Disclinations in crystalline
        solids}, in: {\it Dislocations in Solids Vol.~9}, 
        F.R.N.~Nabarro, ed., North-Holland (1992) p.~191.
\bibitem{GA99} M.Yu.~Gutkin and E.C.~Aifantis, 
    Phys. Stat. Sol. (b)~{\bf 214} (1999) 245.
\bibitem{Mura72} T.~Mura, Arch. Mech.~{\bf 24} (1972) 449.
\bibitem{deWit72b} R.~deWit, Arch. Mech.~{\bf 24} (1972) 499.
\bibitem{Kosseka} E.~Kossecka, Arch. Mech.~{\bf 26} (1974) 995.
\bibitem{VS88} M.C.~Valsakumar and D.~Sahoo, Bull. Mater. Sci.~{\bf 10} (1988) 3.
\bibitem{EL88} D.G.B.~Edelen and D.C.~Lagoudas, {\it Gauge theory and defects in 
        solids}, in: {\it Mechanics and Physics of Discrete System}, Vol.~1,
        G.C.~Sih, ed., North-Holland, Amsterdam (1988).
\bibitem{Maly} C.~Malyshev, Arch. Mech.~{\bf 48} (1996) 1089.
\end{thebibliography}
\end{document}